\documentclass[twocolumn,english,aps,prd,reprint,floatfix,notitlepage,footinbib,preprintnumbers,superscriptaddress]{revtex4-1}
\pdfoutput=1
\usepackage{lmodern}

\usepackage[normalem]{ulem}
\usepackage[T1]{fontenc}
\usepackage[latin9]{inputenc}
\usepackage{geometry}
\usepackage{subfigure,lmodern, amsmath,amssymb, graphicx, pifont, adjustbox, bm, xcolor}
\usepackage{amsfonts}
\usepackage{geometry}
\usepackage{comment}
\usepackage{mathtools}
\usepackage{float}
\usepackage{slashed}
\usepackage{ragged2e}
\usepackage{array}
\usepackage{bbold}
\usepackage{tikz}
\usepackage{tikz-feynman,contour}
\usetikzlibrary{ decorations.markings}
\usetikzlibrary {positioning}
\definecolor {processblue}{cmyk}{0.96,0,0,0}
\usetikzlibrary{arrows.meta}

\makeatletter\g@addto@macro\bfseries{\boldmath}\makeatother

\usepackage{stackengine}
\usepackage{esint}
\usepackage[unicode=true,pdfusetitle,
 bookmarks=true,bookmarksnumbered=false,bookmarksopen=false,
 breaklinks=false,pdfborder={0 0 1},backref=false,colorlinks=true]
 {hyperref}
\hypersetup{
 pdfauthor={Clifford Cheung},
 citecolor=black,linkcolor=black,urlcolor=black}

\newcommand{\appendixref}[1]{\hyperref[#1]{appendix~\ref{#1}}}
\def\equationautorefname~#1\null{eq.\,(#1)\null}
\usepackage{breakurl}
\usepackage{breakurl}
\usepackage[hang,flushmargin]{footmisc} 
\allowdisplaybreaks
\makeatletter

\usepackage{etoolbox}
\apptocmd{\thebibliography}{\justifying\setlength{\leftskip}{7.4mm}}{}{} 
 
 \usepackage{relsize}
\usepackage{babel}

\makeatletter
\def\simgt{\mathrel{\lower2.5pt\vbox{\lineskip=0pt\baselineskip=0pt
           \hbox{$>$}\hbox{$\sim$}}}}
\def\simlt{\mathrel{\lower2.5pt\vbox{\lineskip=0pt\baselineskip=0pt
           \hbox{$<$}\hbox{$\sim$}}}}
\makeatother

\usepackage{changepage}

\newcommand{\be}{\begin{equation}}
\newcommand{\ee}{\end{equation}}
\newcommand{\bea}{\begin{eqnarray}}
\newcommand{\eea}{\end{eqnarray}}

\newcommand{\Eq}[1]{Eq.~(\ref{#1})}

\newcommand{\eq}[2]{\be\begin{aligned}#1 \label{#2}\end{aligned}\ee}

\newcommand{\centertikz}[1]{\vcenter{\hbox{
\begin{tikzpicture} #1 \end{tikzpicture}
}} }

\def\Ablock{X}
\def\Bblock{Y}

\def\laA{\alpha}
\def\laB{\beta}
\def\diag{\text{diag}}
\def\tr{\text{tr}}

\renewcommand{\a}{a}
\renewcommand{\ap}{a'}
\renewcommand{\b}{b}
\newcommand{\bp}{b'}

\newcommand{\dSn}{\Delta S_n(A)}
\newcommand{\Sn}{S_n(A)}
\newcommand{\Snp}{S'_n(A)}

\newcolumntype{P}[1]{>{\centering\arraybackslash}p{#1}}

\begin{document}

\title{On Entropy Growth in Perturbative Scattering}

\author{Clifford Cheung}
\affiliation{Walter Burke Institute for Theoretical Physics, California Institute of Technology, Pasadena, CA 91125}
\author{Temple He}
\affiliation{Walter Burke Institute for Theoretical Physics, California Institute of Technology, Pasadena, CA 91125}
\author{Allic Sivaramakrishnan}
\affiliation{Walter Burke Institute for Theoretical Physics, California Institute of Technology, Pasadena, CA 91125}

\begin{abstract}
\noindent 

Inspired by the second law of thermodynamics, we study the change in subsystem entropy generated by dynamical unitary evolution of a product state in a bipartite system.
Working at leading order in perturbative interactions, we prove that the quantum $n$-Tsallis entropy of a subsystem never decreases, $\Delta S_n \geq 0$, provided that subsystem is initialized as a statistical mixture of states 
of equal probability. This is true for any choice of interactions and any initialization of the complementary subsystem. When this condition on the initial state is violated, it is always possible to explicitly construct a ``Maxwell's demon'' process that decreases the subsystem entropy, $\Delta S_n <0$.  Remarkably, for the case of particle scattering, the circuit diagrams corresponding to $n$-Tsallis entropy are the same as the on-shell diagrams that have appeared in the modern scattering amplitudes program, and $\Delta S_n \geq0$ is intimately related to the nonnegativity of cross-sections.

\end{abstract}
\maketitle

\preprint{CALT-TH 2023-009}

\maketitle

\noindent {\bf Introduction.} The second law of thermodynamics mandates the monotonic growth of disorder in a closed system. It applies whenever the microscopic dynamics have been reformulated in terms of coarse-grained variables. Crucially, even when these macroscopic degrees of freedom are exactly specified, the microstate is not.  We are instead left with an {\it ensemble} of possible microstates consistent with the coarse-grained properties.    Assuming the configuration of the system is suitably generic, as defined by this ensemble, the second law asserts that it will typically evolve into another generic state exhibiting equal or higher thermodynamic entropy.
  
At the same time, it is logically impossible to construct a quantity that depends solely on the microstate of a system and is also nondecreasing in time for {\it any choice of initial microstate and dynamical evolution}. Any such quantity would also be nondecreasing for the time-reversed evolution of the final state, thus yielding a contradiction unless the quantity is trivially constant. However, since the second law pertains to generic states consistent with specified coarse-grained properties, it actually encodes how our ignorance about the details of a system propagates into our future predictions.
  
Meanwhile, in the context of quantum mechanics, subsystem entropy can be derived uniquely from the knowledge of the complete wavefunction of the full system after tracing out subsystem degrees of freedom.  Since the time evolution of the wavefunction is reversible via the Schrodinger equation, subsystem entropy certainly cannot be monotonically increasing for {\it any choice of initial wavefunction and Hamiltonian}.   It is then natural to ask: for which states does subsystem entropy never decrease, independently of the detailed dynamics of the system?

In this paper, we answer this question for a broad class of quantum mechanical systems. In particular, we consider a bipartite system $A\otimes B$ initialized as a product density matrix, $\rho_{AB} = \rho_A \otimes \rho_B$, where $\rho_A$ and $\rho_B$ can in principle be mixed, but have no quantum correlations between them.  Furthermore, we assume that the state evolves by a ``scattering'' process implemented by a unitarity operator $U =1 +iT$ that is perturbatively close to the identity matrix \cite{Note1}.  Throughout our analysis we will compute perturbatively, order by order in the scattering matrix $T$. Despite our focus on scattering, our analysis will also apply to more general quantum systems.

Given that the initial state is a product state, the mutual information of the bipartite system starts at zero and can only grow under unitary evolution. However, the same is {\it not} necessarily true of the subsystem entropy.   Working at leading nontrivial perturbative order, 
we derive a general formula for $\dSn$, the change in the quantum $n$-Tsallis entropy of subsystem $A$.  We then determine the conditions for which it is nonnegative. Because the perhaps more familiar $n$-Renyi entropy is a monotonic function of the $n$-Tsallis entropy, we stress that these conditions also dictate when the leading perturbative $n$-Renyi entropy is nonnegative.

Remarkably, we find that $\dSn \geq 0$ for {\it any choice} for $\rho_B$ and $U$ if and only if $\rho_A$ is proportional to a projection matrix in some basis, which we dub a ``projector state.''   In physical terms, a projector state is a statistical mixture of states which are of exactly equal probability.   
The space of projector states contains as a subset the case where $\rho_A$ is pure, maximally mixed, and everything in between (maximally mixed in a subset of states).    As we will show, if $\rho_A$ deviates even slightly from projector, then there exists a ``Maxwell's demon'' in the form of a $\rho_B$ and $U$ for which $\dSn$ decreases. 
  
A physical interpretation of our result is that there is only one way to guarantee that the entropy of subsystem $A$ is nondecreasing at ${\cal O}(T^2)$, independent of the state of subsystem $B$ and the underlying dynamics defined by $U$.  Subsystem $A$ must be in a statistical mixture in which we are {\it democratically ignorant} of its precise configuration.

Note that all of our results apply to $\dSn$ at its leading nontrivial perturbative order, which is ${\cal O}(T^2)$. In particular, 
we make no claim about higher order corrections to $\dSn$, which can be of either sign.

As a check of our general proof, we have verified that our claims are true in several concrete examples, including a bipartite system of qudits and particle scattering in quantum field theory.  In the latter, we note that quantum $n$-Tsallis entropies are literally equal to the so-called ``on-shell diagrams'' that are commonly studied in the modern amplitudes program \cite{Hodges:2005bf, Hodges:2005aj, Arkani-Hamed:2009hub, Arkani-Hamed:2009ljj, Arkani-Hamed:2009nll, Arkani-Hamed:2012zlh}. Furthermore, when the only quantum numbers labelling the particles are external momenta, we find that $\dSn$ for a projector state is trivially nonnegative since it can be expressed as a positive sum over scattering cross-sections.

\medskip

\noindent {\bf Entropy Diagrams.}   To compute the $n$-Tsallis entropy, we employ a simple circuit diagrammatic formalism for the algebraic manipulation of states and operators. This approach will conveniently reduce the proliferation of indices and their contractions.  Furthermore, as we will discuss later, in the context of quantum field theory, these diagrams are literally the same as the on-shell diagrams.

To begin, let us define the initial density matrix for the $A\otimes B$ system by the diagrammatic element
\eq{
\rho_{AB} = \centertikz{
[decoration={markings, 
    mark= at position 0.5 with {\arrow{stealth reversed}}},
] 
   \node[draw, rounded corners,fill=lightgray, minimum size=0.9cm] (rho) at (0,0) {{\scriptsize $AB$}};
  \draw[postaction={decorate}] (-1,0.2) -- ($(rho.west)+(0,0.2cm)$) node[midway, above] {\tiny $A$};
  \draw[postaction={decorate}] (-1,-0.2) -- ($(rho.west)+(0,-0.2cm)$)  node[midway, below] {\tiny $B$};
  \draw[postaction={decorate}] ($(rho.east)+(0,0.2cm)$) -- (1,0.2) node[midway, above] {\tiny $A$};
  \draw[postaction={decorate}] ($(rho.east)+(0,-0.2cm)$) -- (1,-0.2) node[midway, below] {\tiny $B$};
} \, ,
}{}
where the lines with outgoing and incoming arrows denote the uncontracted indices of kets and bras, respectively.
The reduced density matrix for subsystem $A$ is
\eq{
\rho_A = {\rm tr}_B (\rho_{AB}) = \hspace{-1cm} \centertikz{
[decoration={markings, 
    mark= at position 0.5 with {\arrow{stealth reversed}}},
] 
   \node[draw, rounded corners,fill=lightgray, minimum size=0.9cm] (rho) at (0,0) {{\footnotesize $AB$}};
  \draw[postaction={decorate}] (-1,0.2) -- ($(rho.west)+(0,0.2cm)$) node[midway, above] {\tiny $A$};
  \draw[postaction={decorate}, out=0, in=0] (rho.east) .. controls (2,-1) and (-2,-1)  .. (rho.west) node[midway,below] {\tiny $B$};
  \draw[postaction={decorate}] ($(rho.east)+(0,0.2cm)$) -- (1,0.2) node[midway, above] {\tiny $A$};
 }  \hspace{-1cm}  \, ,
}{}
while concatenating a pair of $\rho_A$ diagrams and taking the trace yields the initial purity of subsystem $A$,
 \eq{
{\rm tr}_A(\rho_A^2) &= \hspace{-1cm}   \centertikz{
[decoration={markings, 
    mark= at position 0.5 with {\arrow{stealth reversed}}},
] 

   \node[draw, rounded corners,fill=lightgray, minimum size=0.9cm] (rho1) at (0,0) {{\footnotesize $AB$}};
   \node[draw, rounded corners,fill=lightgray, minimum size=0.9cm] (rho2) at (2,0) {{\footnotesize $AB$}};
  \draw[postaction={decorate}, looseness=4, out=45, in=45] (rho1.east) .. controls (2,-1) and (-2,-1)  .. (rho1.west) node[midway,below] {\tiny $B$};
  \draw[postaction={decorate}, looseness=4, out=45, in=45] (rho2.east) .. controls (4,-1) and (0,-1)  .. (rho2.west) node[midway,below] {\tiny $B$};
  \draw[postaction={decorate}] ($(rho1.east)+(0,0.2cm)$) -- ($(rho2.west)+(0,0.2cm)$) node[midway, above] {\tiny $A$};
  \draw[postaction={decorate}] ($(rho2.east)+(0,0.2cm)$) .. controls (4,1) and (-2,1)  .. ($(rho1.west)+(0,0.2cm)$) node[midway, above] {\tiny $A$};
 } \hspace{-1cm}   \, .
}{purity_A}
Next, we define the final state density matrix produced by evolution under a unitary $U$, so
\eq{
\rho_{AB}' = U\rho_{AB}U^\dagger  = \centertikz{
 [decoration={markings, 
    mark= at position 0.5 with {\arrow{stealth reversed}}},
] 
   \node[draw, rounded corners,fill=lightgray, minimum size=0.9cm] (rho) at (0,0) {{\footnotesize $AB$}};
   \node[draw, circle, minimum size=1cm] (S) at (1.5,0) {{\footnotesize  $\phantom{{}_\dag}U^\dag$}};
   \node[draw, circle, minimum size=1cm] (Sdag) at (-1.5,0) {{\footnotesize $U$ }};
   \draw[postaction={decorate}] ($(Sdag.east)+(0,0.2cm)$) -- ($(rho.west)+(0,0.2cm)$) node[midway, above] {\tiny $A$};
   \draw[postaction={decorate}] ($(Sdag.east)+(0,-0.2cm)$) -- ($(rho.west)+(0,-0.2cm)$) node[midway, below] {\tiny $B$};
   \draw[postaction={decorate}] ($(rho.east)+(0,0.2cm)$) -- ($(S.west)+(0,0.2cm)$) node[midway, above] {\tiny $A$};
   \draw[postaction={decorate}] ($(rho.east)+(0,-0.2cm)$) -- ($(S.west)+(0,-0.2cm)$) node[midway, below] {\tiny $B$};
  \draw[postaction={decorate}] (-2.5,0.2) -- ($(Sdag.west)+(0,0.2cm)$) node[midway, above] {\tiny $A$};
  \draw[postaction={decorate}] (-2.5,-0.2) -- ($(Sdag.west)+(0,-0.2cm)$)  node[midway, below] {\tiny $B$};
  \draw[postaction={decorate}] ($(S.east)+(0,0.2cm)$) -- (2.5,0.2) node[midway, above] {\tiny $A$};
  \draw[postaction={decorate}] ($(S.east)+(0,-0.2cm)$) -- (2.5,-0.2) node[midway, below] {\tiny $B$};
} \, .
}{rho'_AB}
To compute the final purity of subsystem $A$, we simply
send $\rho_{AB}$ to $\rho'_{AB}$ in the diagram in \Eq{purity_A}, yielding
 \eq{
 {\rm tr}_A(\rho_A^{\prime 2}) &= \hspace{-1.5cm}  \centertikz{
[decoration={markings, mark= at position 0.5 with {\arrow{stealth reversed}}}] 

   \node[draw, rounded corners,fill=lightgray, minimum size=0.9cm] (rho1) at (0,0) {{\footnotesize $AB$}};
   \node[draw, circle, minimum size=1cm] (S1) at (1.5,0) {{\footnotesize $ \phantom{{}_\dag}U^\dag$}};
   \node[draw, circle, minimum size=1cm] (Sdag1) at (-1.5,0) {{\footnotesize $U$}};
   \draw[postaction={decorate}] ($(Sdag1.east)+(0,0.2cm)$) -- ($(rho1.west)+(0,0.2cm)$) node[midway, above] {\tiny $A$};
   \draw[postaction={decorate}] ($(Sdag1.east)+(0,-0.2cm)$) -- ($(rho1.west)+(0,-0.2cm)$) node[midway, below] {\tiny $B$};
   \draw[postaction={decorate}] ($(rho1.east)+(0,0.2cm)$) -- ($(S1.west)+(0,0.2cm)$) node[midway, above] {\tiny $A$};
   \draw[postaction={decorate}] ($(rho1.east)+(0,-0.2cm)$) -- ($(S1.west)+(0,-0.2cm)$) node[midway, below] {\tiny $B$};
  \draw[postaction={decorate}] ($(S1.east)+(0,-0.2cm)$) .. controls (4,-1) and (-4,-1)  .. ($(Sdag1.west)+(0,-0.2cm)$) node[midway, below] {\tiny $B$};
  
  \node[draw, rounded corners,fill=lightgray, minimum size=0.9cm] (rho2) at (0,2) {{\footnotesize $AB$}};
   \node[draw, circle, minimum size=1cm] (S2) at (-1.5,2) {{\footnotesize $ \phantom{{}_\dag}U^\dag$}};
   \node[draw, circle, minimum size=1cm] (Sdag2) at (1.5,2) {{\footnotesize $U$}};
   \draw[postaction={decorate}] ($(Sdag2.west)+(0,-0.2cm)$) -- ($(rho2.east)+(0,-0.2cm)$) node[midway, below] {\tiny $A$};
   \draw[postaction={decorate}] ($(Sdag2.west)+(0,0.2cm)$) -- ($(rho2.east)+(0,0.2cm)$) node[midway, above] {\tiny $B$};
   \draw[postaction={decorate}] ($(rho2.west)+(0,-0.2cm)$) -- ($(S2.east)+(0,-0.2cm)$) node[midway, below] {\tiny $A$};
   \draw[postaction={decorate}] ($(rho2.west)+(0,0.2cm)$) -- ($(S2.east)+(0,0.2cm)$) node[midway, above] {\tiny $B$};
  \draw[postaction={decorate}] ($(S2.west)+(0,0.2cm)$) .. controls (-4,3) and (4,3)  .. ($(Sdag2.east)+(0,0.2cm)$) node[midway, above] {\tiny $B$};
  
  \draw[postaction={decorate}] ($(S2.west)+(0,-0.2cm)$) .. controls (-2.5,1.5) and (-2.5,0.5) .. ($(Sdag1.west)+(0,0.2cm)$) node[midway, right] {\tiny $A$};
  \draw[postaction={decorate}] ($(S1.east)+(0,0.2cm)$) .. controls (2.5,0.5) and (2.5,1.5)  .. ($(Sdag2.east)+(0,-0.2cm)$) node[midway, left] {\tiny $A$};
 } \hspace{-1.5cm}  \, .
}{purity'_A}
By contracting more and more complicated diagrams of this type we can mechanically compute any quantity built from traces of products of density matrices.

The quantum $n$-Tsallis entropies for subsystem $A$ in the initial and final state are
$\Sn = \tfrac{1}{n-1}(1-{\rm tr}_A(\rho_A^n) )$ and $ \Snp  = \tfrac{1}{n-1}(1-{\rm tr}_A(\rho_A^{\prime n}) )$.  For the entirety of our analysis, we will assume that $n\geq 2$ is an integer. The change in $n$-Tsallis entropy is then
\eq{
\dSn &= \Snp-\Sn \\
	&= \tfrac{1}{n-1}  ({\rm tr}_A(\rho_A^n) -{\rm tr}_A(\rho_A^{\prime n})) \,.
}{dS_def}
At this point we have made no assumptions about the initial state density matrix or the unitary evolution. 

Next, we introduce two additional and nontrivial assumptions.  First, we assume that the initial state is a product density matrix,
\eq{
\rho_{AB} = \rho_A \otimes \rho_B = \centertikz{
[decoration={markings, 
    mark= at position 0.5 with {\arrow{stealth reversed}}},
] 
   \node[draw, rounded corners,fill=lightgray, minimum size=0.4cm] (rhoA) at (0,0.3) {{\footnotesize $A$}};
   \node[draw, rounded corners,fill=lightgray, minimum size=0.4cm] (rhoB) at (0,-0.3) {{\footnotesize $B$}};
  \draw[postaction={decorate}] (-1,0.3) -- (rhoA.west) node[midway, above] {\tiny $A$};
  \draw[postaction={decorate}] (-1,-0.3) -- (rhoB.west) node[midway, below] {\tiny $B$};
  \draw[postaction={decorate}] (rhoA.east) -- (1,0.3) node[midway, above] {\tiny $A$};
  \draw[postaction={decorate}] (rhoB.east) -- (1,-0.3) node[midway, below] {\tiny $B$};} \,,
}{rho_AB_prod}
indicating that we start with no quantum correlation between the subsystems.
Second, we assume that the scattering matrix $T$, defined by
\eq{
U = 1+ iT  = \centertikz{
 [decoration={markings, 
    mark= at position 0.5 with {\arrow{stealth reversed}}}] 
   \draw[postaction={decorate}] (0,0.2) -- (1,0.2) node[midway, above] {\tiny $A$};
   \draw[postaction={decorate}] (0,-0.2) -- (1,-0.2) node[midway, below] {\tiny $B$};
}
+
\centertikz{
 [decoration={markings, 
    mark= at position 0.5 with {\arrow{stealth reversed}}},
] 
   \node[draw, circle, minimum size=1cm] (S) at (0,0) {{\footnotesize $iT$ }};
   \draw[postaction={decorate}] ($(S.east)+(0,0.2cm)$) -- (1,0.2) node[midway, above] {\tiny $A$};
   \draw[postaction={decorate}] ($(S.east)+(0,-0.2cm)$) -- (1,-0.2) node[midway, below] {\tiny $B$};
  \draw[postaction={decorate}] (-1,0.2) -- ($(S.west)+(0,0.2cm)$) node[midway, above] {\tiny $A$};
  \draw[postaction={decorate}] (-1,-0.2) -- ($(S.west)+(0,-0.2cm)$)  node[midway, below] {\tiny $B$};
}\, ,
}{U_as_T}
is proportional to a perturbative coupling constant.  For our analysis we will work to ${\cal O}(T^2)$, so it will be important to enforce unitarity via the the optical theorem,
\eq{
i(T-T^\dagger) = -T T^\dagger \, .
}{opt_thm}
Substituting \Eq{rho_AB_prod} and \Eq{U_as_T} into \Eq{dS_def}, we obtain a general formula for the quantum $n$-Tsallis entropy at ${\cal O}(T^2)$ in the perturbative coupling,
\eq{
\dSn &=	 \tfrac{n}{n-1}\times
\centertikz{
[decoration={markings, mark= at position 0.5 with {\arrow{stealth reversed}}}] 
       \node[draw, rounded corners,fill=lightgray, minimum size=0.9cm] (alpha) at (0,1) {{\footnotesize $\Ablock$}};
       \node[draw, rounded corners,fill=lightgray, minimum size=0.9cm] (beta) at (0,-1) {{\footnotesize $\Bblock$}};
       \node[draw, circle, minimum size=1cm] (Sdag1) at (1.5,0) {{\footnotesize $T$}};
   \node[draw, circle, minimum size=1cm] (S2) at (-1.5,0) {{\footnotesize $T$}};
   \draw[postaction={decorate}] ($(alpha.east)+(0,0.2)$) .. controls (0.6,1.2) and (1.3,1) .. (Sdag1.north) node[midway, above] {\tiny $A$};
      \draw[postaction={decorate}] (Sdag1.north west) .. controls (1,0.6) and (0.6,0.8) .. ($(alpha.east)+(0,-0.2cm)$) node[midway, below] {\tiny $A \;$};
      \draw[postaction={decorate}] ($(alpha.west)+(0,-0.2cm)$) .. controls (-0.6,0.8)  and (-1,0.6) .. (S2.north east) node[midway, below] {\tiny $\;A$};
    \draw[postaction={decorate}] (S2.north)  .. controls (-1.3,1)  and (-0.6,1.2) .. ($(alpha.west)+(0,0.2cm)$) node[midway, above] {\tiny $A$};
   \draw[postaction={decorate}] (Sdag1.south) .. controls (1.3,-1) and (0.6,-1.2) .. ($(beta.east)+(0,-0.2cm)$) node[midway, below] {\tiny $B$};
   \draw[postaction={decorate}] ($(beta.east)+(0,0.2cm)$) .. controls (0.6,-0.8) and (1,-0.6) .. (Sdag1.south west) node[midway, above] {\tiny $B\;$};
   \draw[postaction={decorate}] ($(beta.west)+(0,-0.2cm)$) .. controls  (-0.6,-1.2) and (-1.3,-1) .. (S2.south) node[midway, below] {\tiny $B$};
  \draw[postaction={decorate}] (S2.south east) .. controls  (-1,-0.6) and (-0.6,-0.8) .. ($(beta.west)+(0,0.2cm)$) node[midway, above] {\tiny $\;B$};
} \\ \\
\centertikz{
[decoration={markings, mark= at position 0.5 with {\arrow{stealth reversed}}}] 
	\node[draw, rounded corners,fill=lightgray, minimum size=0.9cm] (alpha) at (0,0) {{\footnotesize $\Ablock$}};
	\draw[postaction={decorate}] ($(alpha.east)+(0,0.2)$) -- (1,0.2) node[midway, above] {\tiny $A$};
	\draw[postaction={decorate}] (1,-0.2) -- ($(alpha.east)+(0,-0.2)$) node[midway, below] {\tiny $A$};
	\draw[postaction={decorate}] (-1,0.2) -- ($(alpha.west)+(0,0.2)$)  node[midway, above] {\tiny $A$};
	\draw[postaction={decorate}] ($(alpha.west)+(0,-0.2)$) -- (-1,-0.2)  node[midway, below] {\tiny $A$};
}
&=  \vcenter{
\hbox{\raisebox{-16.5ex}{
\begin{tikzpicture}
[decoration={markings, mark= at position 0.55 with {\arrow{stealth}}}] 
	\node[draw, rounded corners,fill=lightgray, minimum size=0.4cm] (rho1) at (-.7,-0.3) {{\footnotesize $A$}};
	\node[minimum size=0.5cm] (dots) at (0,-0.3) {\footnotesize$\cdots$};
	\node[draw, rounded corners,fill=lightgray, minimum size=0.4cm] (rho2) at (.7,-0.3) {{\footnotesize $A$}};
	\draw[postaction={decorate}] (1.25,0.3) -- (-1.2,0.3) node[midway, above] {};
	\draw[postaction={decorate}] (-1.25,-0.3) -- (rho1.west) node[midway, below] {};
	\draw[] (rho1.east) -- (dots.west) node[midway, below] {};
	\draw[] (dots.east) -- (rho2.west) node[midway, below] {};
	\draw[postaction={decorate}] (rho2.east) -- (1.25,-0.3) node[midway, below] {};
	\draw [thick, decoration={brace, mirror, raise=0.5cm}, decorate] (rho1.west) -- (rho2.east) 
node [pos=0.5,anchor=north,yshift=-0.5cm] {{\footnotesize $n$ copies}}; 
\end{tikzpicture}
}}}
-
 \vcenter{
\hbox{\raisebox{5.5ex}{
\begin{tikzpicture}
[decoration={markings, mark= at position 0.55 with {\arrow{stealth reversed}}}] 
	\node[draw, rounded corners,fill=lightgray, minimum size=0.4cm] (rho2) at (-.7,0.3) {{\footnotesize $A$}};
	\node[minimum size=0.5cm] (dots) at (0,0.3) {\footnotesize$\cdots$};
	\node[draw, rounded corners,fill=lightgray, minimum size=0.4cm] (rho3) at (.7,0.3) {{\footnotesize $A$}};
	\node[draw, rounded corners,fill=lightgray, minimum size=0.4cm] (rho1) at (0,-0.3) {{\footnotesize $A$}};
	\draw[postaction={decorate}] (-1.25,0.3) -- (rho2.west) node[midway, above] {};
	\draw[] (rho2.east) -- (dots.west) node[midway, above] {};
	\draw[] (dots.east) -- (rho3.west)  node[midway, above] {};
	\draw[postaction={decorate}] (rho3.east) -- (1.25,0.3)  node[midway, above] {};
	\draw[postaction={decorate}] (1.25,-0.3) -- (rho1.east) node[midway, below] {};
	\draw[postaction={decorate}] (rho1.west) -- (-1.25,-0.3)  node[midway, below] {};
	\draw [thick, decoration={brace, raise=0.5cm}, decorate] (rho2.west) -- (rho3.east) 
node [pos=0.5,anchor=south,yshift= 0.5cm] {{\footnotesize $n-1$ copies}}; 
\end{tikzpicture}
}}}
\\ 
\centertikz{
[decoration={markings, mark= at position 0.5 with {\arrow{stealth reversed}}}] 
	\node[draw, rounded corners,fill=lightgray, minimum size=0.9cm] (beta) at (0,0) {{\footnotesize $\Bblock$}};
	\draw[postaction={decorate}] ($(beta.east)+(0,0.2)$) -- (1,0.2) node[midway, above] {\tiny $B$};
	\draw[postaction={decorate}] (1,-0.2) -- ($(beta.east)+(0,-0.2)$) node[midway, below] {\tiny $B$};
	\draw[postaction={decorate}] (-1,0.2) -- ($(beta.west)+(0,0.2)$)  node[midway, above] {\tiny $B$};
	\draw[postaction={decorate}] ($(beta.west)+(0,-0.2)$) -- (-1,-0.2)  node[midway, below] {\tiny $B$};
}
&= 
\centertikz{
[decoration={markings, mark= at position 0.5 with {\arrow{stealth reversed}}}] 
	\node[draw, rounded corners,fill=lightgray, minimum size=0.4cm] (rho2) at (1.7,-0.3) {{\footnotesize $B$}};
	\draw[postaction={decorate}] (rho2.west) -- (0.8,-0.3) node[midway, below] {};
	\draw[postaction={decorate}] (2.6,-0.3) -- (rho2.east) node[midway, below] {};
	\draw[postaction={decorate}] (0.8,0.3) -- (2.6,0.3) node[midway, above] {};
}
-
\centertikz{
[decoration={markings, mark= at position 0.5 with {\arrow{stealth reversed}}}] 
	\node[draw, rounded corners,fill=lightgray, minimum size=0.4cm] (rho1) at (-0.4,0) {{\footnotesize $B$}};
	\node[draw, rounded corners,fill=lightgray, minimum size=0.4cm] (rho2) at (0.4,0) {{\footnotesize $B$}};
	\draw[postaction={decorate}] (rho1.south) .. controls (-0.6,-0.6) .. (-0.8,-0.7) node[midway, below] {\tiny $\;B$};
	\draw[postaction={decorate}] (-0.8,0.7) .. controls (-0.6,0.6) .. (rho1.north) node[midway, above] {\tiny $\;B$};
	\draw[postaction={decorate}] (0.8,-0.7) .. controls (0.6,-0.6) .. (rho2.south)  node[midway, below] {\tiny $B \;$};
	\draw[postaction={decorate}] (rho2.north) .. controls (0.6,0.6) .. (0.8,0.7)  node[midway, above] {\tiny $B \;$};
} \,,
}{dS_final}
where in the derivation we have dropped the diagrams where a single $B$ loops on itself, since each such diagram represents the multiplicative factor $\tr_B(\rho_B) = 1$.

Let us elaborate briefly on the explicit derivation of \Eq{dS_final}.
To clarify our discussion, it will be convenient to split the diagrams in \Eq{dS_final} according to their first and second terms, so $X=X_1 - X_2$ and $Y=Y_1-Y_2$. Examining the diagrammatic representation of $\dSn$, we see that $X_1 Y_1$ corresponds to contributions that enter linearly in $T$ or $T^\dagger$.  Since these terms all enter through the combination $T-T^\dagger$, they can be rewritten in terms of $T T^\dagger$ by the optical theorem in \Eq{opt_thm}.  Meanwhile, $X_2 Y_1$ comes from $T T^\dagger$ contributions in which a single $T$ and a single $T^\dagger$ arise from the same factor of $\rho_A'$.  Last but not least, $X_1 Y_2$ and $X_2 Y_2$ come from terms of the form $T^2$, $T^{\dagger 2}$, and $T T^\dagger$, where each $T$ or $T^\dagger$ originates from a distinct factor of $\rho_A'$.  Crucially, since we are only working to ${\cal O}(T^2)$, \Eq{opt_thm} implies that $T$ is Hermitian up to higher order corrections.  This means to our order of interest, we can set $T\sim T^\dagger$ with impunity.

\medskip

\noindent {\bf Proof of Claim.} Let us now derive the conditions under which $\dSn \geq0$ at ${\cal O}(T^2)$.
To begin, we note that since the initial density matrix is a product state, \Eq{rho_AB_prod}, then without loss of generality we can choose a basis that simultaneously diagonalizes both $\rho_A$ and $\rho_B$,
\eq{
(\rho_A)_{\a \ap} = \laA_{\a} \delta_{\a \ap} \qquad \textrm{and}\qquad
(\rho_B)_{\b \bp} = \laB_{\b} \delta_{\b \bp}  \, ,
}{}
where $0\leq \laA_{\a},\laB_{\b}\leq 1$ and $\sum_{\a} \laA_{\a} =\sum_{\b} \laB_{\b} =1$.
Translating \Eq{dS_final} into explicit index notation, we obtain
\eq{ \dSn &= \tfrac{n}{n-1}  \sum_{\substack{\a,\ap}} \laA_{\a} ( \laA_{\a}^{n-1} - \laA_{\ap}^{n-1}) \Gamma_{\a \ap}\\
\Gamma_{\a \ap} &= \sum_{\b,\bp} \Big[  \laB_{\b} |T_{\a \ap \b \bp}|^2 - \laB_{\b} \laB_{\bp} T^{*}_{\a \ap \b \b}T_{\a \ap \bp \bp}  \Big]\,.
}{dS_final_index}
Here we have used the fact that because we are working to ${\cal O}(T^2)$, we can treat $T$ as effectively Hermitian, so $T_{\a\ap \b\bp} = T^{*}_{\ap \a \bp \b}$. 
Meanwhile, $\Gamma_{\a \ap}$ is automatically nonnegative since it can be written as a sum of squares,
\eq{
	&\Gamma_{\a \ap}  = \sum_{\b} \laB_{\b} \Big[ (1-\laB_{\b}) |T_{\a \ap \b \b}|^2 \\
	&\qquad \qquad \qquad + \sum_{\bp \neq \b} \big( |T_{\a \ap \b \bp}|^2 - \laB_{\bp} T^{*}_{\a \ap \b \b}T_{\a \ap\bp \bp} \big) \Big] \\
	&= \sum_{\b} \sum_{\bp \neq \b} \Big[ \laB_\b |T_{\a \ap \b \bp}|^2 + \tfrac12 \laB_{\b} \laB_{\bp} |T_{\a \ap\b\b} - T_{\a \ap \bp \bp}|^2 \Big] \geq 0 \,,
}{betablock}
where we have used the fact that $\sum_b \laB_b = 1$.

We now derive the necessary and sufficient conditions on $\rho_A$ such that $\dSn \geq 0$ for any choice of $T$ and $\rho_B$.   In particular, this is achieved if $\rho_A$ is a projector state,
\eq{
\rho_A =  \diag( \alpha,\alpha,\ldots,\alpha  , 0 ,0 , \ldots ,0) \, ,
}{projector_condition}
where every diagonal entry of $\rho_A$ is either zero or the same nonzero number $\alpha$.  Note that there always exists a basis in which the projector state is written in the form given in \Eq{projector_condition}. If \Eq{projector_condition} holds then $\laA_{\a} ( \laA_{\a}^{n-1} - \laA_{\ap}^{n-1}) \geq 0$, in which case \Eq{dS_final_index} and \Eq{betablock} imply that $\dSn \geq 0$. Thus \Eq{projector_condition} is a sufficient condition.

To show that \Eq{projector_condition} is also necessary,
 it suffices to show that when it is not satisfied, there always exists a choice for $T$ and $\rho_B$ corresponding to a ``Maxwell's demon'' system that produces $\dSn < 0$. In particular, if  \Eq{projector_condition} does not hold, then at least two nonzero diagonal entries of $\rho_A$ are not equal. We can take $\laA_1 < \laA_2$ without loss of generality.  Let us now choose $T_{1212} = T_{2121} = 1$ with all other couplings vanishing, together with $\rho_B = \diag(1,0,0,\ldots,0)$ being a pure state.  Then according to \Eq{dS_final_index}, we find that
$\dSn = \tfrac{n}{n-1} \laA_1 (\laA_1^{n-1} - \laA_2^{n-1}) < 0$. This proves that \Eq{projector_condition} is also a necessary condition to mandate $\dSn \geq 0$ for any choice of $T$ and $\rho_B$, and proves the claim stated in our introductory remarks.

Interestingly, if we assume that $\Gamma_{\a \ap}$ is symmetric, then $\dSn\geq0$ is ensured for arbitrary $\rho_A$, and so  \Eq{projector_condition} is not required.  In this case, \Eq{dS_final_index} becomes
\eq{
\dSn 
&=  \tfrac12 \tfrac{n}{n-1} \sum_{\substack{\a,\ap}} (\laA_{\a} -  \laA_{\ap})(\laA_{\a}^{n-1} -  \laA_{\ap}^{n-1}) \Gamma_{\a \ap}  \geq 0 \, .
}{}
Meanwhile, from \Eq{dS_final_index} we see that 
\eq{
\Gamma_{\a \ap} -\Gamma_{\ap \a} &= \sum_{\b,\bp} (\laB_{\b}-\laB_{\bp}) | T_{\a \ap \b\bp}|^2 \, ,
}{Gamma_diff}
where we used that at ${\cal O}(T^2)$ the scattering matrix is Hermitian.  It is straightforward to construct simple scenarios in which $\Gamma_{\a \ap}$ is automatically symmetric.  

For example, \Eq{Gamma_diff} is zero if the scattering matrix is itself symmetric up to a phase,  $T_{\a \ap \b \bp} =T_{\ap \a \b \bp} e^{i\theta}$, corresponding to a symmetry under swapping the in and out states of the $A$ system.  In the context of a relativistic quantum field theory, this might arise if the $A$ particle is a charge-neutral, time-reversal invariant degree of freedom.  Another example is when $\rho_B$ describes a maximally mixed state, in which case all of its entries are equal, so $\laB_{\b}-\laB_{\bp}=0$ for all $b,b'$ in \Eq{Gamma_diff}.

Last but not least, we emphasize that all of the above results strictly apply to Tsallis entropy for integer $n\geq2$. While it is tempting to try to analytically continue our expressions via $n\rightarrow 1$ to derive the von Neumann entropy, this is not permitted here because our particular diagrammatic approach implicitly assumes a regular series expansion in the coupling constant whose leading term is ${\cal O}(T^2)$.  Since von Neumann entropy involves logarithms of the density matrix, its series expansion includes nonanalytic terms of the form ${\cal O}(T^2 \log T)$.

\medskip

\noindent {\bf Examples.} Our results apply directly to any quantum system which undergoes perturbative unitary evolution.  Recently, there have been a surge of interest in the generation of entanglement from dynamical processes, including black hole scattering \cite{Aoude:2020mlg}, perturbative particle scattering \cite{Balasubramanian:2011wt, Seki:2014cgq, Peschanski:2016hgk, Ratzel:2016qhg, Carney:2016tcs, Faleiro:2016lsf, Fan:2017hcd, Fan:2017mth, Fan:2021qfd, Liu:2022grf}, infrared radiation \cite{Carney:2017jut, Carney:2017oxp}, and nuclear physics \cite{Illa:2022zgu}.  However, these works were not concerned with the specific question of under what conditions will the subsystem entropy increase or decrease. In this section, we consider the implications of our results for two examples: a finite-dimensional system of coupled qudits, and particle scattering in quantum field theory.

\smallskip

\noindent {\it Qudit System.} As a concrete application, let us consider a system where $A \otimes B$ is a pair of coupled qudits.  Such a system arises naturally in the context of scattering if we post-select on the external momenta of particles, keeping track only of internal or spin labels \cite{Aoude:2020mlg}. 

For simplicity, we will first consider the case of qubits before generalizing to qudits. The initial product density matrix is defined in \Eq{rho_AB_prod}, with
\eq{
\rho_{A} = 
\left(
\begin{array}{cc}
\alpha &0 \\
0 & 1-\alpha \\
\end{array}
\right) \quad \textrm{and}\quad
\rho_B =
\left(
\begin{array}{cc}
\beta &0 \\
0 & 1-\beta \\
\end{array}
\right)
\, ,
}{} 
in a diagonal basis where $0\leq \alpha, \beta \leq 1/2$ without loss of generality. Next, we define an arbitrary unitary evolution operator, $U = \exp(i\lambda_{\mu\nu} \sigma^\mu \otimes \sigma^\nu) $, where $\sigma^0 = \mathbb{1}$ and $\sigma^{i}$ for $i=1,2,3$ are the Pauli matrices and $\lambda_{\mu\nu}$ is a four-by-four real matrix of perturbative couplings.

A short calculation yields the change in linear entropy,
\eq{
\Delta S_2(A) &= \tfrac12 \lambda_i K_{ij}(\alpha,\beta)\lambda_j + {\cal O}(\lambda^3) \, ,
}{}
where $\lambda_i = (\lambda_{11},\lambda_{12},\lambda_{13},\lambda_{21},\lambda_{22},\lambda_{23})$ are the six nonzero parameters defining the perturbative dynamics, which we consider up to ${\cal O}(\lambda^2)$. The six-by-six matrix $ K_{ij}(\alpha,\beta)$ has three distinct eigenvalues,
\eq{
&8(1-2\alpha)(\beta-\alpha) \\
&8(1-2\alpha)(1-\alpha-\beta) \\
&16(1-2\alpha)^2(1-\beta)\beta \, .
}{}
If we demand that $\Delta S_2(A) \geq 0$ for any choice of couplings, then these eigenvalues must be nonnegative, so
\eq{
  \alpha=\tfrac12 \qquad  \textrm{or} \qquad \alpha\leq \beta\, .
}{alpha_sol}
We have checked that enforcing $\dSn \geq 0$ for $n\geq 2$ imposes the same conditions.

When the first condition in \Eq{alpha_sol} is satisfied, $\rho_A$ is maximally mixed, so \Eq{dS_final_index} implies that $\dSn=0$ at leading perturbative order.  The higher order contributions to $\dSn$ can only be negative, since $A$ is maximally mixed.  
Meanwhile, the second condition \Eq{alpha_sol} is a mutual relation between $\rho_A$ and $\rho_B$.  If we demand that $\dSn\geq 0$ for {\it any choice} of $\rho_B$, then this includes the case $\beta=0$, in which case $\alpha=0$, so $\rho_A$ is a pure state.   

We have also considered a bipartite system of three- or four-level states, and verified that $\dSn\geq 0$ is ensured at ${\cal O}(\lambda^2)$ for arbitrary $U$ and $\rho_B$ if and only if $\rho_A$ is proportional to a projector.  For example, in the three-level system this corresponds to $\rho_A =\textrm{diag}(1,0,0), \textrm{diag}(\tfrac12,\tfrac12,0)$, or $\textrm{diag}(\tfrac13,\tfrac13,\tfrac13)$.  
Thus, an important difference from the two-level system is that there now exists a choice for $\rho_A$ which is neither pure not maximally mixed.   In the special case that $\rho_A$ is maximally mixed, the entropy inequality is always saturated, at least at leading perturbative order.

Note that the inequality $\dSn\geq 0$ will not be saturated in general.  However, when it is saturated, the corrections at higher order in perturbation theory can be negative \cite{Note2}.

\smallskip
\noindent {\it Particle Scattering.}  Our results also apply to particle scattering in relativistic quantum field theory.   In this case, the space of states is infinite-dimensional and parameterized by external momenta.  As we will see, in this context the change in the quantum $n$-Tsallis entropy and its nonnegativity follow directly from the fact that differential cross-sections are nonnegative.

For example, consider a quantum field theory describing a pair of perturbatively interacting scalar fields, $A$ and $B$.   As in the finite-dimensional case, we assume without loss of generality that the initial density matrices are diagonal, so $\rho_A =  \int_p \alpha_p |p\rangle \langle p|$ and $\rho_B =  \int_q \beta_q |q\rangle \langle q|$, where $\int_p= \int \frac{d^3 p}{2\omega_{p}}$ and $\omega_{p}$ is the free particle energy corresponding to momentum $p$, and similarly for $q$. Here, $\int d^3p \, \alpha_p=\int d^3q \, \beta_q=1$, and our normalization for the density matrices implies that ${\rm tr}_A( \rho_A) = {\rm tr}_B( \rho_B) = \delta^3(0) =V$, which is equal to the formally divergent volume of space. Meanwhile, the unitary evolution is controlled by the scattering matrix $T$, which encodes the amplitude for elastic scattering between $A$ and $B$.  

To compute $\dSn$, we can use our diagrammatic formula in \Eq{dS_final}, properly generalized to the case of particle scattering.  Concretely,  the $A$ and $B$ lines denote the Fock space identity operators,
\eq{
\centertikz{
 [decoration={markings, 
    mark= at position 0.5 with {\arrow{stealth reversed}}}] 
   \draw[postaction={decorate}] (0,0.2) -- (1,0.2) node[midway, above] {\tiny $A$} node[midway, below] {\phantom{\tiny $A$}};
}
= \int\limits_p |p\rangle \langle p| \qquad \textrm{and} \qquad
\centertikz{
 [decoration={markings, 
    mark= at position 0.5 with {\arrow{stealth reversed}}}] 
   \draw[postaction={decorate}] (0,0.2) -- (1,0.2) node[midway, above] {\tiny $B$} node[midway, below] {\phantom{\tiny $B$}};
}
= \int\limits_q |q\rangle \langle q|   \, ,
}{}
where $\langle p| p'\rangle = 2\omega_p \delta^3(p-p')$, and likewise for $q$.   Furthermore, we note that closed loops of a single $B$ particle now yield the normalization factor $ {\rm tr}_B( \rho_B) =V$, which for finite-dimensional systems would normally be one.  Thus, we should modify the $\Bblock$ diagram in  \Eq{dS_final} by multiplying the first term by $V^{n-1}$ and the second term by $V^{n-2}$.   After we properly include these factors,  \Eq{dS_final_index} becomes
\eq{
	\dSn &= \tfrac{n}{n-1} \int\limits_{p,p'} \alpha_p(\alpha_p^{n-1} - \alpha_{p'}^{n-1}) \Gamma_{p,p'} \\
	\Gamma_{p,p'} &= V^{n-2} \int\limits_{q,q'} \Big[
	 \beta_q |T_{p,q\rightarrow p',q'}|^2  
	V \\
	&\qquad \qquad \quad \;\;   -   \beta_q \beta_{q'} T^*_{p,q\rightarrow p' q}T_{p,q'\rightarrow p' q'}\Big]\,,
}{dS_QFT}
where $T_{p,q\rightarrow p',q'} = \delta^4(p+q-p'-q') A_{p,q\rightarrow p',q'}$ only has support on kinematic configurations that conserve energy and momentum.
As a result, the second term in $\Gamma_{p,p'}$ is proportional to $\delta^4(p-p')$, so it zeros out when inserted back into $\dSn$
\cite{Note3}.

In summary, the change in subsystem entropy is 
\eq{
\dSn&=	 \tfrac{n}{n-1} V^{n-1}\!\!\!\!\!\! \int\limits_{\substack{p,p',q,q'}}\!\!\!\!\!\! \alpha_p(\alpha_p^{n-1}  - \alpha_{p'}^{n-1})  \beta_q |T_{p,q\rightarrow p',q'}|^2 \, .
}{dS_QFT_alt}
Since the differential cross-section, $ |T_{p,q\rightarrow p',q'}|^2$, is nonnegative, we can straightforwardly deduce the conditions under which $\dSn$ is nonnegative.  In particular, the analog of the necessary and sufficient condition for $\dSn \geq 0$ in \Eq{projector_condition} is that $\alpha_p \subseteq (0,\alpha)$ is a function whose only values are either zero or a single positive constant. Assuming this condition, we see that $\alpha_p(\alpha_p^{n-1} - \alpha_{p'}^{n-1}) $ in  \Eq{dS_QFT_alt}  is nonnegative.  Consequently, $\dSn$ is literally a nonnegative integral over the differential cross-section, and hence automatically nonnegative.  Second, exactly following our general analysis above, we see that if $\beta_q$ is a constant, or if the scattering matrix satisfies $T_{p,q\rightarrow p',q'}=T_{p',q\rightarrow p,q'} e^{i\theta}$, then $\dSn \geq 0$ for any choice of $\alpha_p$.

Amusingly, a corollary of our above discussion is that the diagram in \Eq{dS_final} is literally equal to sum of multi-loop, perturbative Feynman diagrams with all legs localized onto unitary cuts.  Such objects, known as on-shell diagrams, have appeared ubiquitously in the study of generalized unitarity \cite{Eden:1966dnq, Cachazo:2008vp, Bern:2011qt}, which is a highly efficient approach to building perturbative higher-loop integrands from tree amplitudes.  Furthermore, on-shell diagrams have appeared in the study of gauge theories in twistor space \cite{Hodges:2005bf,Hodges:2005aj, Arkani-Hamed:2009hub,Arkani-Hamed:2009ljj, Arkani-Hamed:2009nll} and have been instrumental in new formulations of scattering amplitudes as volumes of polyhedra \cite{Arkani-Hamed:2013jha,Arkani-Hamed:2013kca}.  In all of these applications, the utility of on-shell diagrams follows from the fact that all off-shell redundancy is eliminated by the unitarity cuts, which is also why $\dSn$ is a gauge invariant quantity.

\medskip

\noindent {\bf Discussion.}  We have calculated the perturbative change in $n$-Tsallis entropy $\dSn$, 
generated by a unitary operator $U=1+iT$ acting on 
a product state $\rho_{AB}=\rho_A\otimes\rho_B$. Working to ${\cal O}(T^2)$ in the scattering matrix, we have proven that $\dSn \geq0$  for {\it any choice of $U$ and $\rho_B$} if and only if $\rho_A$ is initialized in a statistical mixture of states whose nonzero probabilities are all equal.  For even the slightest perturbation away from this choice, there exist $U$ and $\rho_B$ that will decrease the subsystem entropy.

We emphasize that our results were obtained without making any detailed assumptions beyond the perturbativity of $U$, and that our setup encapsulates an enormous range of phenomena, including all scattering processes at weak coupling. However, the technical manipulations in this work rely heavily on the assumption that our initial state is a product state, and it would be interesting to see if there exists a wider class of states where $\dSn \geq 0$ under suitable conditions. Imposing restrictions on $U$ or $\rho_B$ may broaden the set of allowed $\rho_A$ as well.

Our analysis leaves several promising avenues for future inquiry. First, while we have treated $U$ as an operator characterizing dynamical unitary evolution in time, it can be interpreted far more broadly.  In particular, it would be interesting to consider systems in which $U$ instead describes an adiabatic perturbation relating two {\it different physical systems}.  For example, consider a system with a product ground state, $\rho_A \otimes \rho_B$.  If we now add a perturbative and adiabatically small correction to the dynamics, the ground state shifts to $\rho_{AB}'$.  Our calculation of $\dSn$ also characterizes the difference in ground state subsystem entropy in these two scenarios.

Second, it would be interesting to explore explicit physical scenarios in which nature realizes a product initial state. For instance, this will happen for any setup where two subsystems are initially decoupled but then brought into causal contact. A joining quench is one well-studied example of this process \cite{Eisler_2007,Calabrese:2007mtj}. As more speculative examples, one might imagine a system of black holes or baby universes which is initialized as a product state before becoming entangled through some dynamical process \cite{Giddings:1987cg,Coleman:1988tj, Maldacena:2013xja}. It would be intriguing to explore the possible ramifications, if any, our results might have in these very different scenarios.

\vspace{3mm}

\noindent {\it Acknowledgments:}  C.C.,~T.H.,~and A.S.~are all supported by the Department of Energy (Grant No.~DE-SC0011632) and by the Walter Burke Institute for Theoretical Physics. T.H. and A.S.~are also supported by the Heising-Simons Foundation ``Observational Signatures of Quantum Gravity'' collaboration grant 2021-2817. We are grateful to Ning Bao, Daniel Carney, Soonwon Choi, Julio Parra-Martinez, and Grant Remmen for insightful discussions and useful comments on the draft.

\vspace{-5mm}

\providecommand{\href}[2]{#2}\begingroup\raggedright\endgroup

\end{document}